\documentclass{biophysEDITED}
\usepackage{helvet,times}
\usepackage{bm,textcomp}

\gridframe{N}
\cropmark{N}

\usepackage{amsmath}

 \usepackage{graphicx}

\usepackage{color,soul}
\usepackage{amssymb}
\renewcommand{\emph}[1]{{\itshape #1}}
\usepackage[left,modulo]{lineno} 
\newcommand{\ignore}[1]{}
\twocolumn

\usepackage[round,numbers,sort&compress]{natbib}

\begin{document}

\setcounter{page}{1} 

\title{Phosphorylation-induced mechanical regulation of intrinsically disordered neurofilament protein assemblies}

\author{Eti Malka-Gibor, $^{1,\dagger}$Micha Kornreich, $^{1,\dagger}$Adi Laser-Azogui, $^1$Ofer Doron, $^1$Irena Zingerman-Koladko, $^2$Ohad Medalia, $^3$and Roy Beck$^{1,*}$}

\address{$^1$Raymond and Beverly Sackler School of Physics and Astronomy, Tel Aviv University, Israel \\ $^2$Department of Life Sciences and the National Institute for Biotechnology in the Negev, Ben-Gurion University, Israel $^3$Department of Biochemistry, University of Zurich, Winterthurerstrasse 190, Switzerland}



\begin{abstract}%
{The biological function of protein assemblies was conventionally equated with a unique three-dimensional protein structure and protein-specific interactions. However, in the past 20 years it was found that some assemblies contain long flexible regions that adopt multiple structural conformations. These include neurofilament (NF) proteins that constitute the stress-responsive supportive network of neurons. Herein, we show that NF networks macroscopic properties are tuned by enzymatic regulation of the charge found on the flexible protein regions. The results reveal an enzymatic (phosphorylation) regulation of macroscopic properties such as orientation, stress-response and expansion in flexible protein assemblies. Together with a model explaining the attractive electrostatic interactions induced by enzymatically added charges, we demonstrate that phosphorylation-regulation is far richer and versatile than previously considered.}

\end{abstract}

\maketitle 
\section*{INTRODUCTION}

In the past two decades it was discovered that
approximately 50\% of human proteins are intrinsically disordered, {\itshape i.e.} they
contain long peptide regions that do not fold into secondary or tertiary
structures. The disordered regions contain a disproportionate number
of phosphorylation sites, which are functionally crucial \cite{Uversky2011}.
In particular, assemblies of phosphorylation-rich intrinsically disordered proteins (IDPs) are associated
with neurodegenerative diseases including amyotrophic lateral sclerosis,
Alzheimer's, Parkinson's and Charcot-Marie-Tooth \cite{Chen2005,Stoothoff2005,Liu2011}.
Despite the significance of IDP {\itshape assemblies}, little is known
about the phosphorylation regulation of their structural and mechanical
properties.


\begin{figure*}[t!]
\centering
\includegraphics[width=17.8cm]{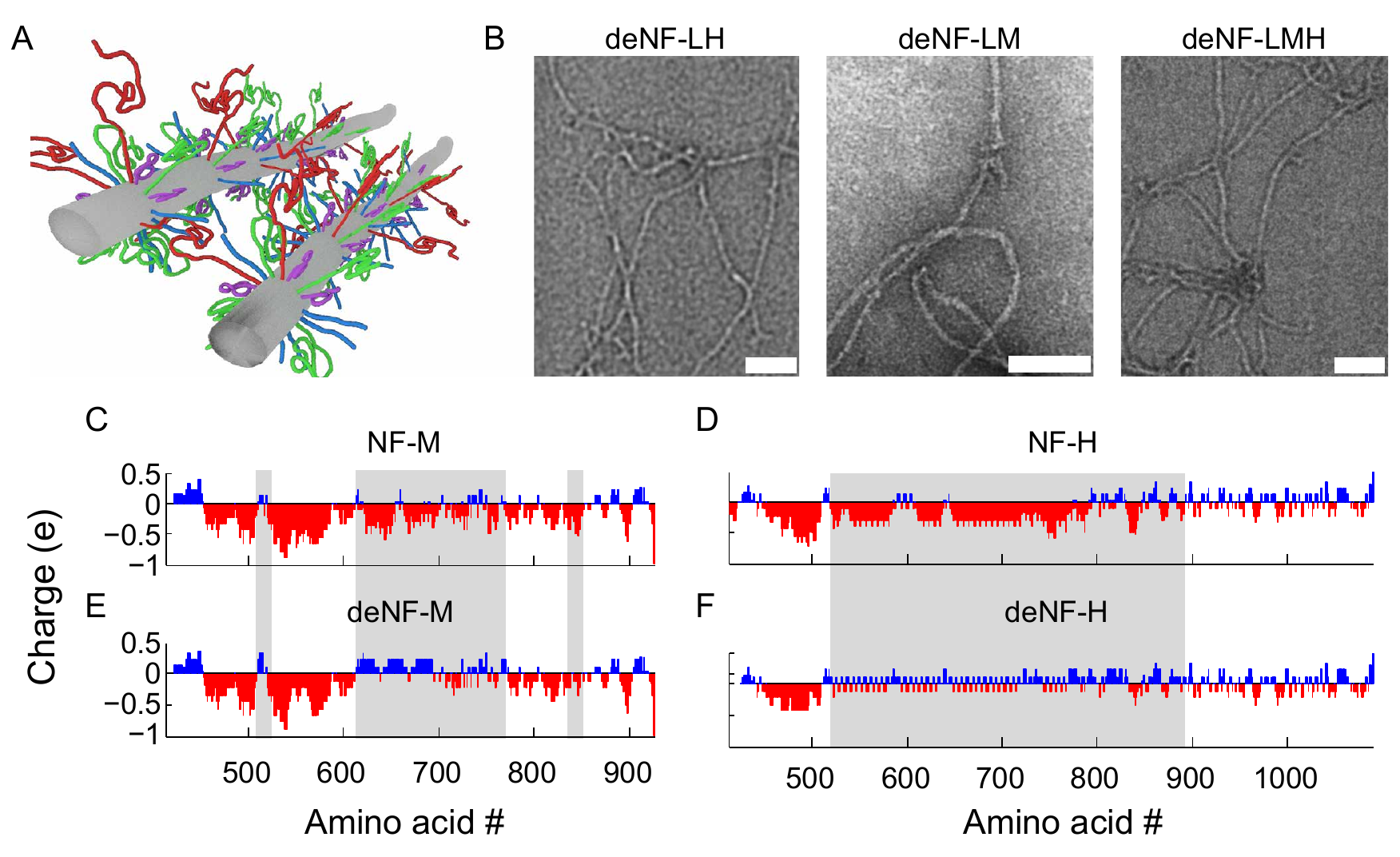}
\caption{Structure of bottlebrush NF filaments. (A) A schematic of two neighboring
interacting filaments. Each filament consists of three different subunit
proteins, whose protruding tails are shown in colour (red, blue and
green). (B) Transmission electron microscopy images of de-phosphorylated
filaments (from left to right): NF-LH, NF-LM and NF-LMH. Scale bar
= 100 nm. (c-f) Tail charge distributions of (C) phosphorylated NF-M,
(D) phosphorylated NF-H, (E) de-phosphorylated NF-M, and (F) de-phosphorylated
NF-H. The charge distributions were calculated at {\itshape p}H$=6.8$
and averaged over a 5-amino acid window (see materials and methods).
The grey boxes highlight the protein segments which are most affected
by the phosphate charge removal.}
\label{fig:charge distributions}
\end{figure*}

NFs make a valuable model system for the exploration of phosphorylation-driven
interactions in IDP assemblies, due to their high modularity in protein
content and phosphorylation levels. In axons, NF proteins hierarchically
form a filamentous network whose main roles are to provide the cell
with its mechanical support and structure \cite{Hirokawa1984}. Protein
subunits co-assemble into 10 nm bottlebrush-like filaments (Fig. \ref{fig:charge distributions}A,B),
where the filament backbone consists of the N-terminal domains. The
long disordered C-terminal tail domains protrude outwards, and mediate
inter-filament interactions and neuronal cytoskeletal organization
\cite{Safinya2015,Sihag2007a}. 

The expression levels of NF proteins are modified during nerve growth
or trauma, where changes in protein ratios are thought to accommodate
the changing mechanical and structural needs of the nervous system.
This modularity is enabled by three of the NF proteins, NF-L, NF-M,
and NF-H, which are expressed in both the central and peripheral nervous
systems \cite{Laser-Azogui2015,Yuan2012b}. Their disordered tails,
which govern the network organization, differ considerably in sequence,
length, net charge and charge distribution (Fig. \ref{fig:charge distributions}C-F
and Table S1).

Additional functional versatility is achieved by enzymatic regulation
of tail charge distribution through phosphorylation and de-phosphorylation.
The majority of the phosphorylation sites correspond to the serine
residues of the Lys-Ser-Pro (KSP) repeat motifs. These repeats abound
in the tail domains of NF-M and NF-H subunits, and significantly alter
their charges \cite{Trimpin2004}. For example, de-phosphorylation
of NF-H tail changes its total charge from -97 to -7 e, and charge
density from -0.14 to -0.01 e/amino acid (Fig. \ref{fig:charge distributions}
and Table S1). Given this significant charge difference,
NF-M and NF-H phosphorylation is thought to govern the lateral extensions
of NF tails, thereby regulating NF spacing, axonal calibre and protein
transport \cite{Dale2012,Sihag2007a,Kriz2000,Ackerley2003,Shea2003}.
These structural roles are the focus of a recent debate, following
a study that found no dependence of axonal caliber on the phosphorylation
of NF-M \cite{Barry2012}. 

Two seemingly contradictory effects can be attributed to the phosphorylation-induced
charge regulation. On the one hand, phosphorylation enhances the electrostatic
repulsion due to its excess charge. On the other hand, phosphorylation
can promote short-range attractive cross-bridging between the disordered
tails \cite{Aranda-espinoza2002,Hisanaga1989,Gou1998,Leermakers2008,Jeong2014a,Eyer1988}.
Here we address the structural and mechanical aspects of these two
opposing phosphorylation effects. We show that NF assemblies have
a distinctive phosphorylation-induced regulation of network properties,
such as compression resistance and orientation. The regulation stems
from the structural plasticity of the disordered tails, and therefore
differentiates between IDP and structured-protein assemblies. 

\section*{MATERIALS AND METHODS}

\subsection*{Calculation of basic sequence-based properties }

Protein sequences are imported from UniProt database \cite{UniProt2016},
and their identifiers are: P02548 (NF-L), O77788 (NF-M) and P19246
(NF-H). Since a UniProt-verified Bovine NF-H sequence is still missing,
the mouse NF-H sequence is used instead. Sequence-based calculations
of NF tail charge are performed at $p$H 6.8 using the EMBOSS amino
acid $p$Ka table \cite{Hancock2004}, and the average amino acid volume
$\left(V_{a}=0.134\,{\rm nm^{3}}\right)$ calculation follows Ref.
\cite{Harpaz1994}. The phosphoserine sites are determined using the
UniProt database, and the phosphoserine $p$Ka$_{2}$ is set to 6.2
for charge calculations. Average tail charges and tail fraction charges
appear in Table S1.

\subsection*{Protein purification }

NF subunits (NF-L, NF-M and NF-H) are purified from bovine spinal
cord using a modification of an earlier protocol \cite{Jones2008,Kas1996}.
Spinal cords are homogenized in an equal volume of buffer A (0.1 M
MES, 1 mM EGTA, 1 mM MgCl$_{2}$, 0.02\% (w/v) sodium azide, 7 mM
$\beta$-mercaptoethanol\emph{, p}H 6.8 with NaOH) with 1\% (w/v)
Triton X-100 and 1 mM phenylmethylsulfonyl fluoride. The homogenates
are centrifuged at 30k RPM (Beckman rotor type 45-Ti) for 70 min at
4$\,^{\circ}$C. An equal volume of glycerol is added to the supernatant
and incubated overnight. A pellet of NFs is recovered from the glycerol
solution by precipitation at 40k RPM (Beckman rotor type 45-Ti) for
90 min at 4$\,^{\circ}$C. The pellet is homogenized in buffer A with
0.8 M sucrose and clarified by spinning through a step gradient of
0.8 M sucrose buffer (0.8 M sucrose in buffer A) layered on top of
1.5 M sucrose buffer (1.5 M sucrose in buffer A) for 4 h at 55k RPM
(Beckman rotor type 70-Ti). The pellet is homogenized in buffer B
(0.1 M potassium phosphate and 0.1\% (v/v) $\beta$-mercaptoethanol
in 8 M urea,\emph{ p}H 6.5), and applied to a DEAE sepharose column
(DEAE Sepharose fast flow column, GE Healthcare). The column is rinsed
with buffer B containing 55 mM NaCl which elutes NF-H and protein
contaminates. The next elution step, performed with buffer B at \emph{p}H
7 containing 200 mM NaCl, elutes NF-L and NF-M. Using hydroxylapatite
(HT) column chromatography (hydroxylapatite bio gel HT, Bio-Rad),
the contaminants are removed from the NF-H fractions. NF-L and NF-M
are separated by HT column with a gradient of 0.1 to 0.4 M potassium
phosphate \emph{p}H 7.0. The purity and separation of the NF-H, NF-M
and NF-L subunit proteins are verified by SDS-PAGE (Fig. S1).

\subsection*{Protein de-phosphorylation}

For de-phosphorylated networks, purified proteins are dialyzed against
de-phosphorylation buffer (50 mM Tris \emph{p}H=8, 100 mM NaCl and
10 mM MgCl$_{2}$), and then incubated with 50 units alkaline phosphatase
(CIP, New England Biolabs) per 0.1 mg at 37~$^{\circ}$C overnight.
The process is monitored by the decreased mobility of de-phosphorylated
NF-M and NF-H in SDS-PAGE \cite{Carden1985,Pant1988} (Fig. S1). 

De-phosphorylation of NF-M is further supported by liquid chromatography\textendash mass
spectrometry, performed at the Biological Services Department, Weizmann institute. Technical details are found in the Supporting Material. 

We compare our results against the listed NF-M phosphosites in the
UniProt database. We identified 7 out of the 22 experimentally verified
NF-M phosphosites, as well as 7 out of 8 of the predicted phosphosites
\cite{UniProt2016}. None of the sites were phosphorylated (Table S2). Notably, since bovine NF-L tails contain only few (1-3)
phosphorylation sites \cite{Trimpin2004}, networks composed of recombinant
and native NF-L proteins do not differ in inter-filament spacing,
orientation or compression response \cite{Kornreich2015}.

\subsection*{Filament self-assembly and hydrogel formation}

Following protein purification and de-phosphorylation, protein subunits
are mixed in denaturing conditions at the desired composition. The
protein solution is dialyzed against a MES buffer (100 mM MES, \emph{$p$}H
6.8, 1 mM EGTA, 1 mM MgCl$_{2}$, 0.02\% (w/v) sodium azide, 7 mM
$\beta$-mercaptoethanol and a total 150 mM of NaCl and NaOH salts)
at 37~$^{\circ}$C for 48 hours. 

For optical microscopy and small-angle X-ray scattering (SAXS), the reassembled filaments solution
is centrifuged for 1 h at 50k RPM using a TLA120.1 rotor in a Beckman
Coulter Optima TLX ultracentrifuge, and the supernatant is immediately
removed from the pellet. The NF pellet is transferred to 1.5 mm quartz
capillaries, overlaid with $\approx$100 $\mu$l MES buffer solution
and sealed with epoxy glue to prevent dehydration. To osmotically
pressurize the network, the overlaying MES buffer solution is supplemented
with 20,000 g/mol polyethylene-glycol (PEG)\cite{Parsegian1986a}.
The resultant PEG-induced osmotic pressure $\Pi$ is determined by
the PEG weight percentage (PEG$_{{wt}}$), and follows the formula
$\log_{10}\Pi=1.57+2.75({\rm PEG}{}_{{wt}})^{0.21}$.

To determine the protein molar ratios in the assembled hydrogel, control
samples are analysed by SDS-PAGE as described in Refs. \cite{Jones2008,Kornreich2015}.
The protein molar ratios of the different hetero-filaments are 4:1
for NF-L:NF-H (denoted NF-LH), 7:3 for NF-L:NF-M (denoted NF-LM) and
10:3:2 for NF- L:NF-M:NF-H (denoted NF-LMH).

\subsection*{Transmission electron microscopy }

For transmission electron microscopy (TEM), a sample of 10 $\mu l$
is laid on a formvar coated 400 mesh grid (\#3440c-FA, SPI Supplies)
and then fixed and negatively stained as in Ref. \cite{Mucke2004}.
Images of filaments composed of de-phosphorylated subunits are shown
in Fig. \ref{fig:charge distributions}B, and resemble images of phosphorylated
protein filaments \cite{Beck2010,Kornreich2015}.

\subsection*{Cross polarising microscopy (CPM)}

The hydrogel orientation (isotropic or birefringent nematic) is characterized
by crossed polarised light microscopy (Fig. \ref{fig:phase-cross}).
Sedimented NFs are observed in 1.5 mm quartz capillaries using a Nikon
Eclipse LV 100 POL microscope fitted with 5-20X objectives. Micrographs
are taken with a Nikon D90 camera.

\subsection*{Small angle X-ray scattering (SAXS)}

NF hydrogels 2D diffraction data was integrated azimuthally and the
intensity was plotted versus reciprocal distance $q$. The intensity,
in arbitrary units, showed a broad peak with a maximum in the range
of $q=0.1-0.2$ ${\rm {nm}^{-1}}$. The peak location relates to the
inter-filament spacing ($d=2\pi/q$). Broadening of this peak is observed
due to density fluctuations and the semi-flexible nature of the individual
filaments. Baseline background of the form $A\cdot q^{-B}+C$ with
$B=2-3$ is subtracted from the integrated data, and the resultant
peak is fitted with a Lorentzian function using Matlab (MathWorks) routines \cite{Beck2010}. 

Preliminary experiments were performed at our home-lab using a Pilatus
300K detector and a Xenocs GeniX Low Divergence CuK$\alpha$ radiation
source setup with scatterless slits \cite{Li2008}. Subsequent measurements
were performed at synchrotron facilities: I22 beamline in Diamond,
England; SWING beamline in SOLEIL, France; and I911 SAXS beamline
in MAX-lab, Sweden with 10 keV.

\subsection*{Bulk modulus calculations}

For these calculations, we treat filament as infinitely long impenetrable
cylinders of radius $R_{\mathrm{cyl}}=5$ nm, set in a hexagonal lattice
(Fig. S3 and Ref. \cite{Lab2020}). We define a prism-shaped
unit cell whose base is an equilateral triangle with side length $d$,
which is the inter-filament distance. The prism height is $l=$45
nm, which is the NF protein's rod length. As 32 tails emanate from
each filament backbone every 45 nm \cite{Herrmann2004}, the unit cell
contains a total of $32/6\cdot3=16$ tails. The surface of the
equilateral triangle found in the hexagonal model is $S\left(d\right)=\sqrt{3}d^{2}/4-\pi R_{\mathrm{cyl}}^{2}/2$
, and therefore the unit cell volume holds $V\left(d\right)=S\left(d\right)\cdot45$
nm. The volume fraction is given by $\phi=NV_{a}/V$, where $N$ is
the number of tail amino acids in $V$. The $\Pi$ vs. $d$ compression
curves are fitted with smoothing splines (Fig. \ref{fig:saxs pd}),
and then used for the $B_{T}$ calculation:
\begin{equation}
B_{T}=-V\frac{\mbox{d}\Pi}{\mbox{d}V}=-S\left(d\right)d\frac{\mbox{d}\Pi}{\mbox{d}d}\left(\frac{\mbox{d}V\left(d\right)}{\mbox{d}d}\right)^{-1} \label{eq:BulkMod}
\end{equation}

\section*{RESULTS AND DISCUSSION}
We investigate the mechanical and structural roles of tail phosphorylation by reconstituting filaments from purified NF protein subunit, at desired subunit compositions. To produce de-phosphorylated
filaments, the natively phosphorylated proteins are enzymatically
treated with alkaline phosphatase. Composite filaments include
NF-L with either NF-M (NF-LM), NF-H (NF-LH) or both (NF-LMH). 10 nm wide filament formation is verified by transmission electron
microscopy (TEM, Fig. \ref{fig:charge distributions}B), as reported
for native NFs \cite{Beck2010a,Hisanaga1989}. For simplicity, we
refer to the de-phosphorylated filaments by a ``de'' prefix ({\itshape i.e.},
deNF-LM, deNF-LH or deNF-LMH). 

\begin{figure}
\centering
\includegraphics[width=8.5cm]{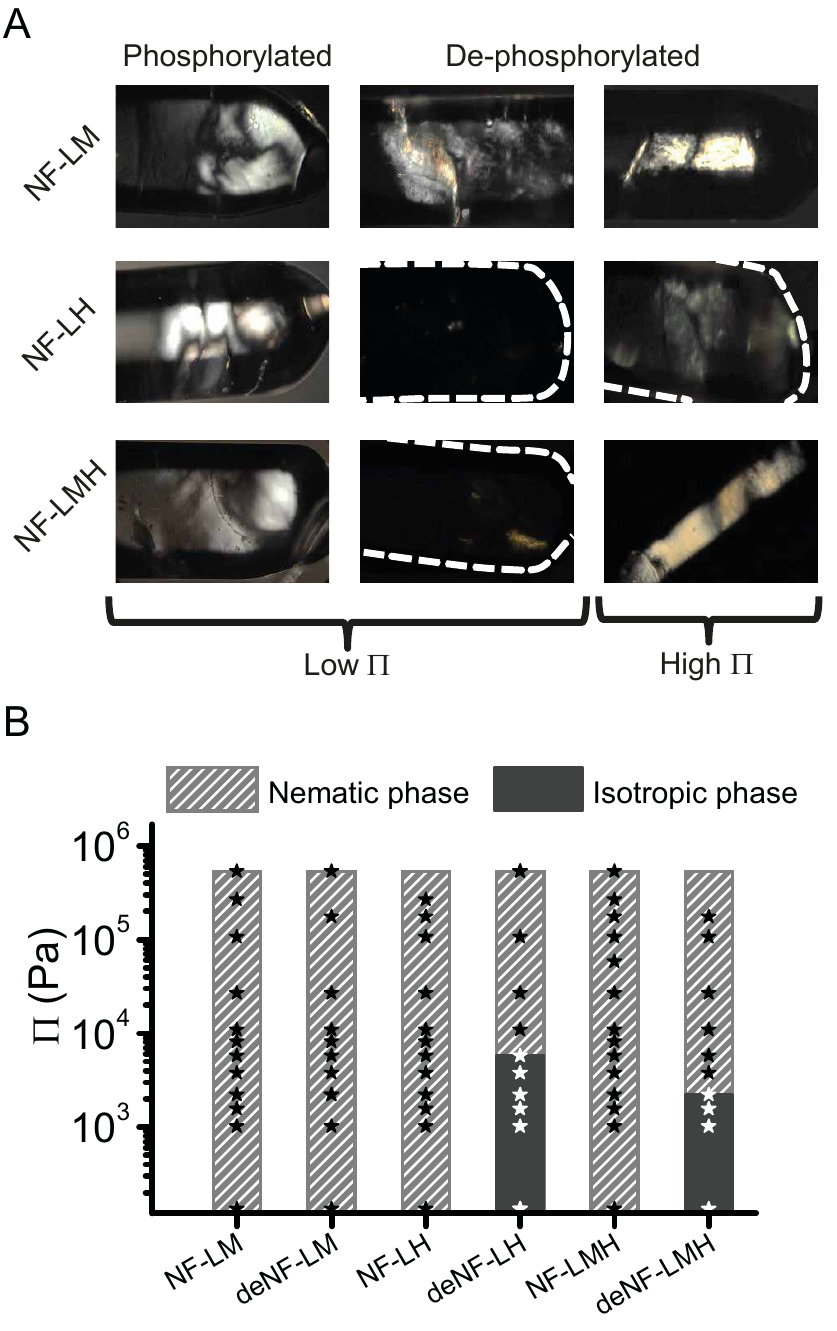}
\caption{NF network phase behaviour, as determined by CPM. (A) Bright field
and CPM images of phosphorylated and de-phosphorylated networks at
low ($10^{2}-10^{3}$ Pa) and high ($10^{5}-2\cdot10^{5}$ Pa) osmotic
pressure in quartz capillaries. White dashed lines demark the capillary
boundaries, as observed with bright field (see Fig.
S2). The filaments in all networks are aligned (nematic) except for
deNF-LM and deNF-LMH, which are isotropic ({\itshape i.e.}, un-oriented)
at low $\Pi$. Each capillary is approximately 1.5 mm wide. (B) Phase
diagram showing the network phase behaviour at different $\Pi$'s.
Each star $\star$ denotes a measurement point.}
\label{fig:phase-cross}
\end{figure}

\begin{figure*}
\centering
\includegraphics[width=17.8cm]{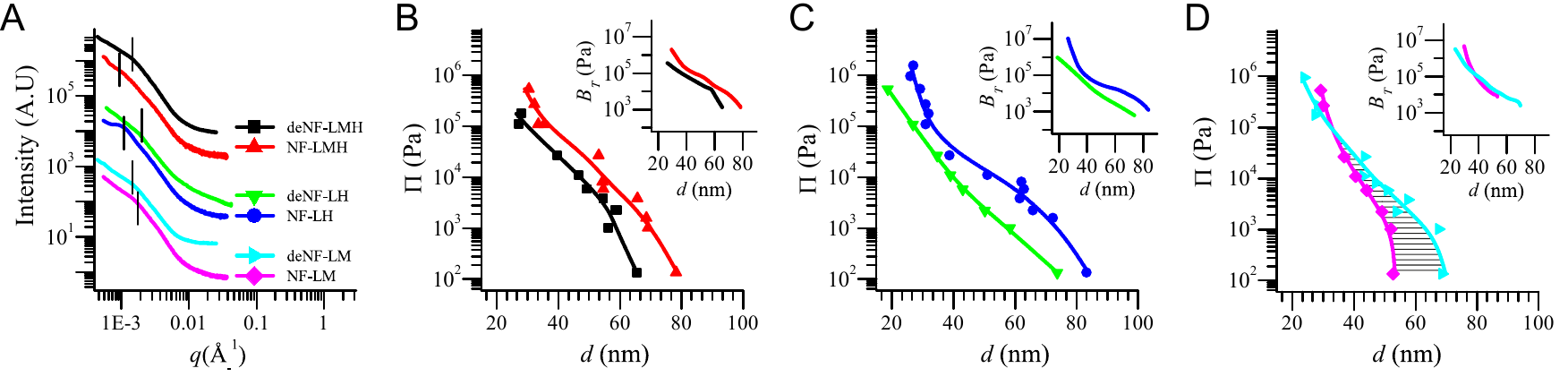}
\caption{Comparison of native and de-phosphorylated networks using SAXS and
osmotic pressure. (A) Intensity curves of NF-LH, NF-LM, and NF-LMH
native and de-phosphorylated networks at 1\% (w/w) PEG ($\Pi=2.2\cdot10^{3}$
Pa). (b-d) Semi-log plot of osmotic pressure $\Pi$ vs. the inter-filament
distance $d$ of different network compositions: (B) NF-LMH, (C) NF-LH
and (D) NF-LM. For $\epsilon$ calculation in Eq. \ref{eq:Epsilon},
we integrate over the textured area in (D). Osmotic bulk modulus $B_{T}$
is shown in insets. The data is fitted with smoothing splines, which
were then used to calculate $B_{T}$. Subunit molar ratios for native
and de-phosphorylated NF-LM, NF-LH and NF-LMH filaments are 7:3 (NF-L:NF-M),
4:1 (NF-L:NF-H) and 10:3:2 (NF-L:NF-M:NF-H). }
\label{fig:saxs pd}
\end{figure*}

At high concentrations, filaments condense into a hydrogel network
that phase separates from the supernatant. To characterize the mechanical
response of the network, we vary the osmotic pressure ($\Pi$) using
polyethylene-glycol (PEG) osmolyte. All phosphorylated filaments as
well as deNF-LM filaments form large oriented (nematic) domains, either
at low or high osmotic pressures, as determined by cross-polarised
microscopy (CPM). In contrast, deNF-LH and deNF-LMH filaments form
less oriented hydrogels with smaller nematic domains (Fig. \ref{fig:phase-cross}).
At low osmotic pressure, deNF-LH and deNF-LMH are isotropic and transition
to a nematic phase at elevated osmotic pressure. Nonetheless, in both
cases, even at elevated osmotic pressures, the aligned domains still
appear smaller and less illuminated than those observed for native
and deNF-LM hydrogels at comparable osmotic pressures. Hence, NF-H
phosphorylation regulates the macroscopic orientation of the hydrogel
networks. Furthermore, a comparison of the CPM images at various osmotic
pressures indicates that deNF-M promotes orientational order while
deNF-H hinders it (Fig. \ref{fig:phase-cross}). 

To study the nanoscopic structural organization and mechanics of the
hydrogel, we measure the inter-filament spacing, $d$, using small-angle
X-ray scattering \cite{Kornreich2015,Beck2010,Kornreich2013}. Azimuthally
averaged intensity curves of phosphorylated ({\itshape i.e.}, native)
and de-phosphorylated networks at $\Pi=2200$ Pa, are shown in Fig.
\ref{fig:saxs pd}A. For each intensity curve, the correlation peak
position ($q_{0}$) is denoted by a vertical line, and is related
to the inter-filament distance by $d=2\pi/q_{0}$. At this given osmotic
pressure, de-phosphorylation of NF-LMH and NF-LH results with a decrease
of the inter-filament spacing from $70$ to $58$ nm and from $67$
to $50$ nm, respectively. Therefore, native phosphorylation of NF-LH
and NF-LMH results with network expansion. This agrees with the conventionally
considered roles of NF phosphorylation, where the substantial addition
of charged phosphate groups is naively expected to stretch the NF
tails and consequently increase the inter-filament spacing. The $\Pi$
vs. $d$ curves of NF-LH and NF-LMH exhibit the same phosphorylation-dependent
effect over a wider range of osmotic pressures (Fig. \ref{fig:saxs pd}B,C).

In contrast, the correlation peak of deNF-LM shifts to lower $q$-values,
in comparison to the NF-LM peak (Fig. \ref{fig:saxs pd}A). This indicates
that the spacing actually increases from $48$ to $60$ nm due to
de-phosphorylation. The result is atypical of charged polymers, as
de-phosphorylation reduces the net charge by $\sim50\%$, and thus
the electrostatic repulsion between adjacent filaments is expected
to decrease. This behaviour indicates an attractive interaction between
phosphorylated tails, which opposes the trend observed in NF-LH and
NF-LMH networks (Fig. \ref{fig:saxs pd}B,C). Notably, the expansion
of deNF-LM in comparison to NF-LM is reversed only at high osmotic
pressures, $\Pi\gtrsim10^{5}$ Pa (Fig. \ref{fig:saxs pd}D).

\begin{figure*}[t!]
\centering
\includegraphics[width=16cm]{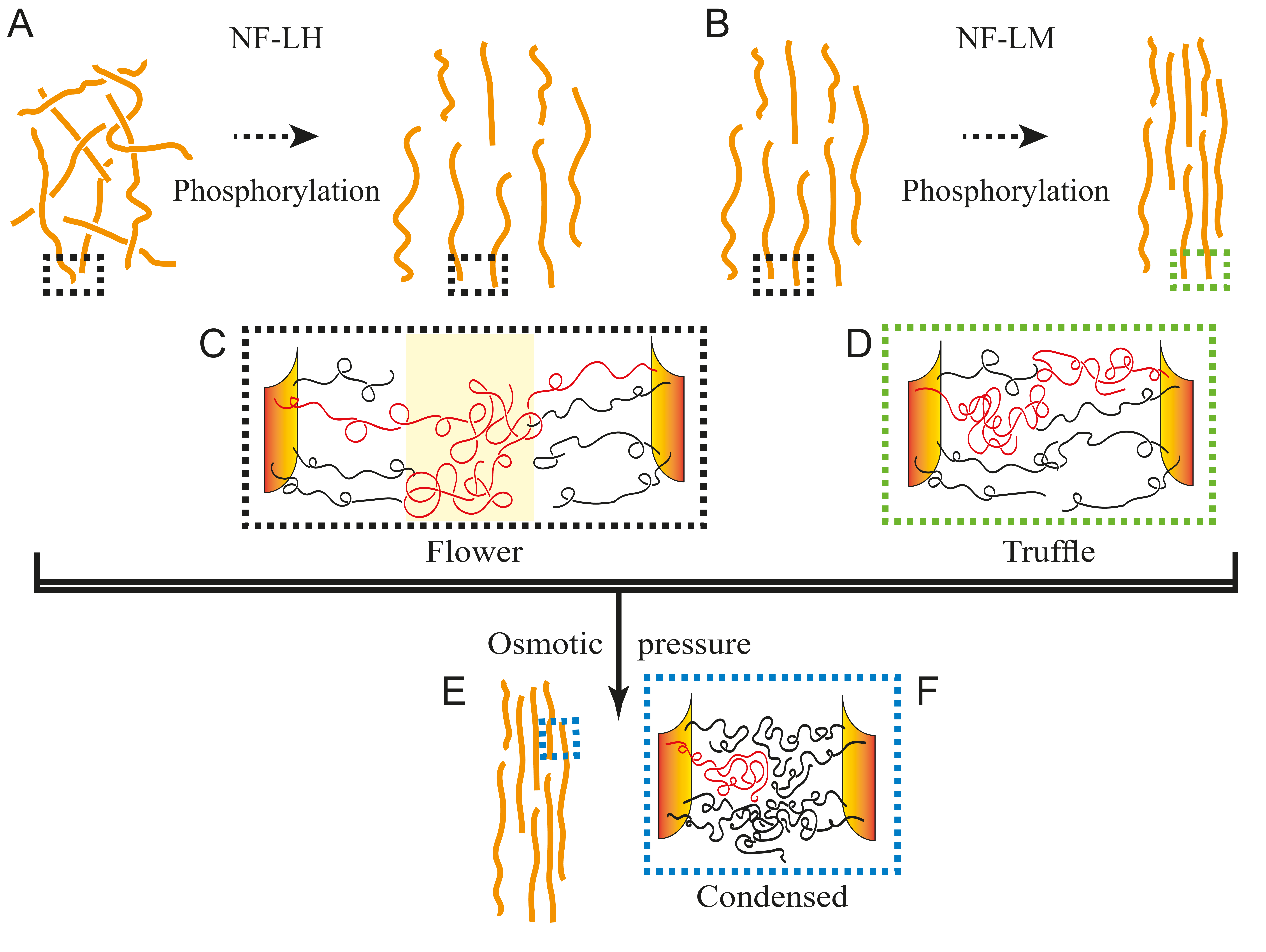}
\caption{Schematic of phosphorylation regulation of NF network expansion, alignment and osmotic stress response. (A) NF-H phosphorlyation aligns the isotropic deNF-LH network and increases the inter-filament distance, whereas (B) NF-M phosphorylation collapses the nematic deNF-LM network. Except for NF-LM networks, all protruding tails organize in two corona layers at low osmotic pressure (C). The outer layer is formed by the long tails (either NF-H or NF-M), and is denoted by a yellow background. Upon phosphorylation, deNF-LM tails transition from a (C) flower to a (D) truffle conformation. (E) Under significant osmotic compression, filaments align and compress (F) while opposite tails increasingly inter-penetrate. }
\label{fig:Schematic}
\end{figure*}

The protein-dependent regulation of network expansion, alignment and osmotic stress response is schematically summarized in Fig. \ref{fig:Schematic}, together with conjectured tails microscopic organization (Fig. \ref{fig:Schematic}C,D,F). Phosphorylation of the NF-H tail aligns the network and increases the inter-filament distance (Fig. \ref{fig:Schematic}A). In contrast, NF-M phosphorylation has little effect on network alignment, and it surprisingly \textit{reduces} the inter-filament distance (Fig. \ref{fig:Schematic}B). This indicates that NF-LM and NF-LH tails are organized differently. 

At low osmotic pressures, tails are expected to form two distinctive layers, known as the flower conformation (Fig. \ref{fig:Schematic}C). The inner layer corona
is composed of the short NF-L tails while NF-H tails are repelled farther away from the corona into the outer layer. Since the tails within
the outer layer are less dense, they assume a flower-like conformation \cite{Leermakers2010a,Kornreich2015,Jayanthi2013}. However, this picture does not agree with the NF-LM tails organization. The inter-filament spacing of NF-LM and NF-L networks are very similar  \cite{Kornreich2015,Beck2010}, suggesting that the long NF-M tails are hidden within the NF-L inner coronas (termed ``truffle'' regime) \cite{Kornreich2015,Lab2020}. Under significant osmotic pressure, all filament types align and compress, while opposite tails increasingly inter-penetrate (Fig. \ref{fig:Schematic}E,F).

In living cells, the osmotic pressure is induced by the crowded environment.
We note that the effect of phosphorylation on the compression response
is much more pronounced than its effect on NF expansion. In particular,
phosphorylation at a given osmotic pressure, does not change the inter-filament
spacing by more than 25\%. However, in order to maintain the same
spacing at the phosphorylated state, the osmotic pressure needs to
increase by up to two orders of magnitudes. 

To further characterize this mechanical response, we calculate the
osmotic bulk modulus, $B_{T}$, which
quantifies the network's resistance to compression (Eq. \ref{eq:BulkMod} and Fig. \ref{fig:saxs pd}B-D,
insets). Similar to the phosphorylation dependent network
expansion and collapse, we find that changes to the mechanical response
are also protein specific. However, the effect of phosphorylation
on the mechanical response is more pronounced, and $B_{T}$ is altered
by as much as an order of magnitude. For NF-LH and NF-LMH, de-phosphorylation
of the tails reduces $B_{T}$ at all $d$. In contrast, de-phosphorylation
of NF-LM increases the network's resistance to compression ({\itshape i.e.},
larger $B_{T}$) at low osmotic pressure. The latter occurs despite
the reduced repulsive electrostatic forces and suggests that the charged
phosphates also take part in attractive interactions \cite{Aranda-espinoza2002,Beck2010}.
Further support to the excess electrostatic repulsion at high and
low osmotic pressures are given by comparison to polymer scaling theories
(Fig. S4).

To estimate the attractive bridging energy per NF-LM phosphosite,
we follow the calculation performed to quantify the attractive
hydration energy between DNA double-helices \cite{Rau1992}. The free
energy is derived from the $\Pi-d$ diagram, under the hexagonal approximation.
To evaluate the average energy per phosphosite ($\epsilon$), we integrate
over the free energy difference between the two phosphorylation states,
and divide by the number of phosphosites in the volume ($N_{p}$):
\begin{equation}
\epsilon=-\frac{k_{\rm B}T}{N_{p}}\int\limits _{0}^{\Pi_{{\rm int}}}\left(V_{{\rm deNF-LM}}-V_{{\rm NF-LM}}\right)\mbox{d}\Pi.\label{eq:Epsilon}
\end{equation}
The integration is performed on the fitted smoothing-spline curves
from $\Pi=0$ to the intersection of the curves at $\Pi_{{\rm int}}\sim10^{5}$
Pa (See textured area in Fig. \ref{fig:saxs pd}D), and yields approximately
$\epsilon=-8\, k_{\rm B}T$ per phosphosite. This is comparable to the
free energy of protein salt bridges \cite{Kumar1999}. Notably, the
integration also includes contributions from repulsive interactions,
and therefore $\epsilon$ provides a lower limit for the attractive
average energy per phosphosite. 

\begin{figure}
\centering
\includegraphics[width=8.5cm]{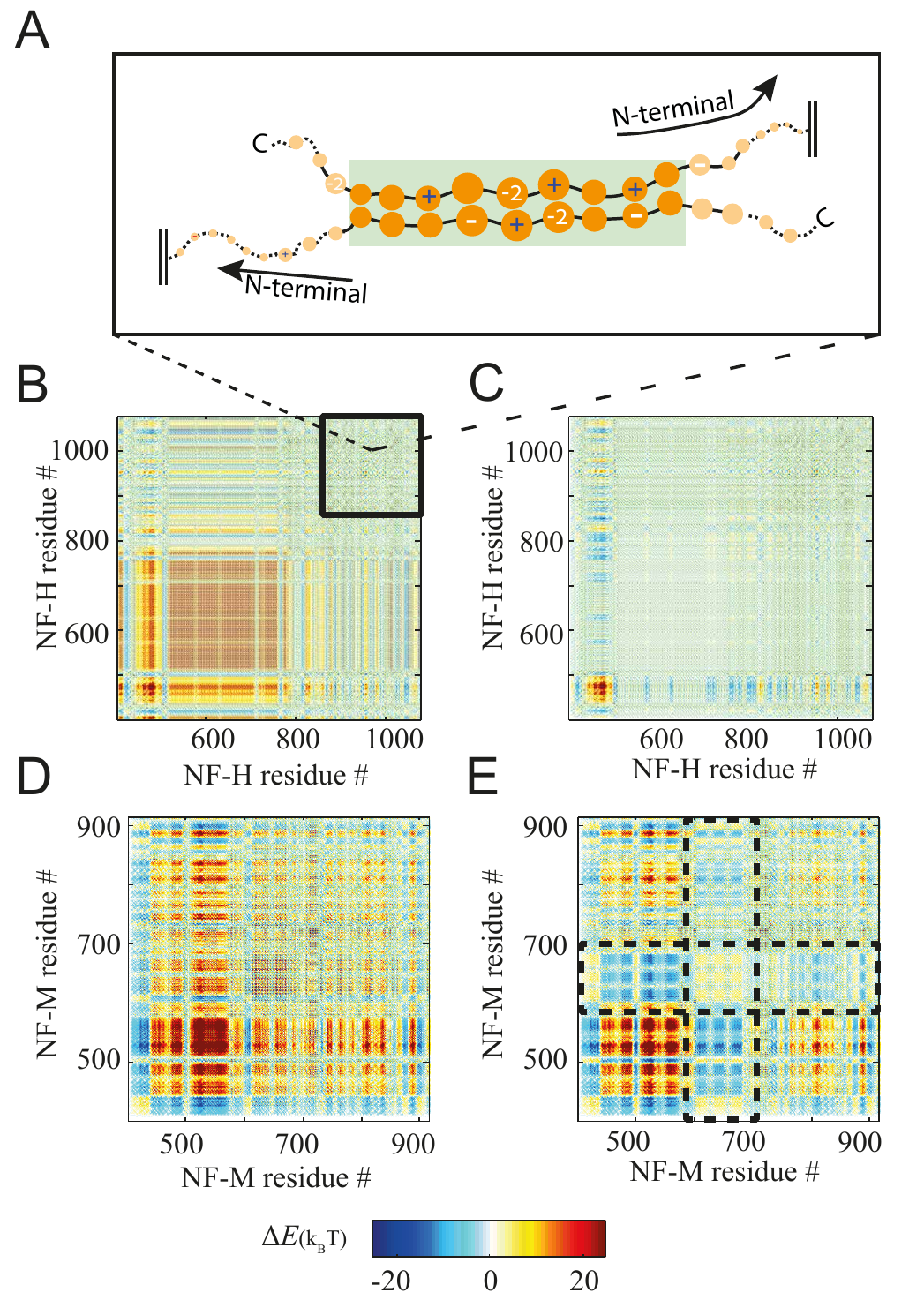}
\caption{Handshake analysis of short-ranged electrostatic interactions between
NF tail regions. (A) Two opposite 9 amino-acid long segments interacting.
Details of the electrostatic interaction energy calculations are found
in Refs. \cite{Beck2010,Kornreich2015}. (b-e) Energy matrices for
all possible segment pairs for two opposing (B) NF-H, (C) deNF-H,
(D) NF-M, or (E) deNF-M tails. Interactions between two oppositely
charged segments, which are more electrostatically viable, are denoted
by blue colours in the interaction matrices. Solid box in (B) marks
the last 200 amino-acids that are known to engage in attractive interactions.
Dashed boxes in (E) denote phosphosite-rich segments (see also Fig.
\ref{fig:charge distributions}). Comparison of (D) with (E) reveals
that de-phosphorylation forms new negative energy pairs between segments
found further away from the filament backbone. }
\label{fig:HandShake}
\end{figure}

Since these attractive interactions are sequence-dependent, they may
account for the opposite phosphorylation expansion trends of NF-LM
and NF-LH. To identify the polypeptide segments involved in such attraction,
we employ a coarse-grained ``handshake'' calculation, aimed at locating
pairs of amino acid segments that interact via electrostatic bridges
\cite{Beck2010}. We calculate the unscreened Coulomb energy of two
interacting segments, where each segment is centered at a specific
tail amino acid and the segment length is on the order of the polypeptide
persistence length (approximately $3$ nm \cite{Bright2001}). We thus obtain a 2D matrix which points at the
most electrostatically viable cross-linking pairs (Fig. \ref{fig:HandShake}A).

Handshake analysis of the natively phosphorylated NF-H tail reveals
multiple potential attractive sites located at its last 200 amino
acids (Fig. \ref{fig:HandShake}B). This was also described by a more
elaborate theoretical calculation \cite{Leermakers2008} and agrees
with previous experiments \cite{Chen2000}. Upon de-phosphorylation,
new attractive sites are predicted near the filament backbone (Fig.
\ref{fig:HandShake}B). These could participate in intra-filament
attractive interactions, in agreement with the de-phosphorylated NF-H
tail collapsed conformation (Figs. \ref{fig:saxs pd}). 

The unexpectedly collapsed conformation of phosphorylated NF-LM may
involve interactions between NF-M segments with either NF-L or NF-M
segments, as \ignore{we have}recently shown \cite{Kornreich2015,Lab2020}.
Here, upon de-phosphorylation, new NF-M attractive binding sites are
formed further away from the filament backbone (Fig. \ref{fig:HandShake}D,E).
Consequently, the loss of excess phosphate charge actually {\itshape expands}
the NF-M tail, as shown in Fig. \ref{fig:saxs pd}D. 

\section*{CONCLUSIONS}

We demonstrate the roles of phosphorylation in regulating the structural
and mechanical properties of NF networks. The phosphorylation-induced
modifications strongly depend on the NF subunit composition, and can
result with either network expansion or collapse (Fig. \ref{fig:Schematic}). This versatility
originates from the dual nature of the induced interactions, which
are both repulsive and attractive and are protein-sequence dependent.

The attractive interactions are clearly manifested in the deNF-LM
network, where the removal of the excess phosphate charges unexpectedly
results with network expansion. This trend relates to previous studies,
which suggested that phosphorylation can promote NF binding either
by associated proteins, by exposing hydrophobic residues, or by direct
involvement in electrostatic bridging \cite{Aranda-espinoza2002,Gou1998,Leermakers2008,Eyer1988}.
Of these possibilities, our results affirm the direct involvement
of NF-M phosphosites in attractive interactions, which is justified
by the stronger ionic bridging formed between the divalent phosphate
group and basic amino acids (Fig. \ref{fig:HandShake}). 

The repulsive interactions govern NF-LH and NF-LMH compression response.
Here, phosphorylation moderately increases the inter-filament distance
and considerably enhances the osmotic compression resistance, characterized
by the bulk modulus (Fig. \ref{fig:saxs pd}). This questions the
hypothesized main structural role of tail phosphorylation, {\itshape i.e.}
to expand the NF network, and suggests a primarily {\itshape mechanical
}role to NF phosphorylation. 

The individual roles of NF-M and NF-H phosphorylation, as demonstrated
here, significantly increase opportunities to regulate NF network
physical properties. Therefore, the roles of NF expansion (Fig. \ref{fig:saxs pd}),
mechanics \cite{Eyer1988}(Fig. \ref{fig:saxs pd}) and
orientation \cite{Deek2013,Storm2015}(Fig. \ref{fig:phase-cross})
must be considered where simultaneous changes in composition and tail
phosphorylation levels are observed. Such changes, for example, occur
during neuronal growth and development, following injury, and in neurodegenerative
diseases \cite{Laser-Azogui2015,Dale2012,Toman2016}. Specifically,
our results indicate a close relation of phosphorylation and NF compression
response and orientation, which was scarcely considered before \cite{Safinya2015}. 

Additional assemblies of hyper-phosphorylated disordered proteins
are also involved in neurodegenerative diseases. These include the
disordered tau and $\alpha$-synuclein proteins, which are hyper-phosphorylated
in pathological inclusions \cite{Chen2005,Stoothoff2005}. Their aggregation
is commonly attributed to an indirect process, where phosphorylation-driven
conformational changes expose new segments for attractive interactions.
However, the electrostatic attraction observed in phosphorylated NF-M
tails demonstrates that phosphates in disordered protein assemblies
{\itshape directly} engage in significant attractive interactions. This
suggests that more attention should be drawn to the role of phosphorylation-driven
attractive electrostatics in the study of disordered assemblies.

\section*{SUPPORTING MATERIAL}

Supplement to this article is available upon request.

\section*{AUTHOR CONTRIBUTIONS}
R.B, A.A.Z, E.M.G, and M.K planned and initiated the project. E.M, M.K,
A.A, O.D, I.K and O.M conducted the experiments. E.M and M.K prepared
the samples and analysed the data. E.M, M.K and R.B wrote the paper.

\section*{ACKNOWLEDGMENTS}
We are grateful to Dr. Geraisy Wassim of Beit Shean abattoirs Tnuva
for kindly providing us with the spinal cords. We thank Ekaterina
Zhulina for useful discussions and suggestions. We thank the following
beamlines for SAXS measurements: I911-SAXS at MAX IV Laboratory, Lund,
Sweden; SWING at SOLEIL synchrotron, Paris, France; and I-22 beamline
at Diamond, England. This work was supported by Israeli Scientific
Foundation (571/11, 550/15) , the Tel Aviv University Center for Nanoscience
and Nanotechnology and the Scakler Institute for Biophysics at Tel
Aviv University. Travel grants to synchrotron facilities were provided
by BioStruct-X. 
\bibliographystyle{biophysj}
\bibliography{library}








\newpage
\onecolumn

\newpage




\end{document}